\documentclass[epsf,twocolumn,showpacs,preprintnumbers]{revtex4}
\usepackage{graphics}
\usepackage{graphicx}% Include figure files
\usepackage{dcolumn} % Align table columns on decimal point
\usepackage{bm}
\usepackage{color}
\usepackage{epsfig}
\newcommand{\bfpo}{BaFe$_2$(PO$_4$)$_2$}
\begin{document}
\title{Large orbital moment and spin-orbit enabled Mott transition\\ 
 in the Ising Fe honeycomb lattice of BaFe$_2$(PO$_4$)$_2$
}
\author{Young-Joon Song$^1$}
\author{Kwan-Woo Lee$^{1,2}$}
\email{mckwan@korea.ac.kr}
\author{Warren E. Pickett$^3$}
\email{pickett@physics.ucdavis.edu} 
\affiliation{
 $^1$Department of Applied Physics, Graduate School, Korea University, Sejong 339-700, Korea\\
 $^2$ Department of Display and Semiconductor Physics, Korea University, Sejong 339-700, Korea \\
$^3$Department of Physics, University of California, Davis,
  CA 95616, USA
}
\date{\today}
\begin{abstract}
\bfpo~ is an unusual Ising insulating ferromagnet based on the Fe$^{2+}$ 
spin $S$ = 2 ion,
the susceptibility of which suggests a large orbital component
to the Fe local moment. 
We apply density functional theory based methods to obtain a microscopic picture
of the competing interactions and the critical role of spin-orbit coupling
(SOC)  in this honeycomb lattice system. 
The low-temperature ferromagnetic phase displays a half-semimetallic Dirac point
pinning the Fermi level and preventing gap opening
before consideration of SOC, presenting a case in which correlation effects
modeled by a repulsive Hubbard $U$ fail to open a gap.
Simultaneous inclusion of both correlation and SOC drives a large orbital moment 
in excess of 0.7 $\mu_B$ (essentially $L$ = 1) for spin aligned 
along the $\hat{c}$ axis,
with a gap comparable with the inferred experimental value.
The large orbital moment accounts for the large Ising anisotropy, 
in spite of the small magnitude
of the SOC strength on the 3$d$ (Fe) ion. 
Ultimately, the Mott-Hubbard gap is enabled by degeneracy lifting
by SOC and the large Fe moments, rather than by standard
Hubbard interactions alone. We suggest that competing orbital occupations
are responsible for the structural transitions involved in the observed re-entrant
rhombohedral-triclinic-rhombohedral sequence. 
\end{abstract}
\maketitle

\begin{figure}[tbp]
%\vskip 8mm
{\resizebox{8cm}{5cm}{\includegraphics{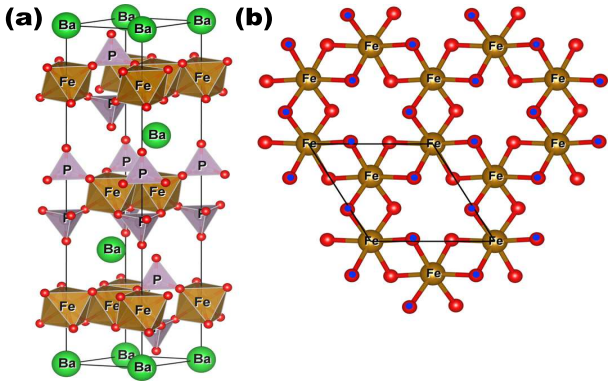}}}
\caption{(Color online) (a) The rhombohedral structure, which contains 
 PO$_4$ tetrahedra and edge-sharing FeO$_6$ octahedra, of \bfpo~ 
 in its low and high $T$ regimes.
 (b) Top view of the two-dimensional Fe honeycomb lattice in the $\hat{a}-\hat{b}$ plane.
 The solid line indicates the unit cell, containing two Fe ions (brown circles). 
 The oxygen ions (small red circles) in the upper layer are denoted by a dot.
% P ions sit the upper and lower layers of the center of the Fe honeycomb lattice.
}
\label{str}
\end{figure}

\section{Introduction}
In the last few decades, the two-dimensional (2D) honeycomb lattice has attracted 
interest due to several exotic physical phenomena related with superconductivity, magnetism,
and topological phases, and to the Dirac point 2D material graphene.  
Recently, discoveries of topological insulators, originally on the honeycomb
lattice,\cite{kane} have stimulated 
increased research on properties of systems with honeycomb 
lattices.\cite{fiete,shitade,zhou} 

Transition-metal $d^1$ systems are good examples to investigate the interplay 
among lattice, spin, orbital degrees of freedom, and correlation effects, 
since variation of $3d\rightarrow 4d\rightarrow 5d$
leads to weakening strength of correlation, but strengthening of spin-orbit coupling (SOC).
For $5d^1$ systems, a large SOC results in
a $J_{eff}=\frac{3}{2}$ Dirac (or relativistic) Mott insulator 
in Ba$_2$NaOsO$_6$\cite{bnoo1,bnoo2} in which the orbital moment plays
a pivotal role.
Similarly, a $J_{eff}=\frac{1}{2}$ Mott transition has been proposed
in $5d^5$ Sr$_2$IrO$_4$, i.e., one hole in the $t_{2g}$ manifold.\cite{jyu}
$4d^1$ systems usually have a reduced SOC and some enhanced correlation strengths.
%  In Nb$_{12}$O$_{29}$,\cite{nb12} a highly {\it localized} $d^1$ Nb$^{5+}$ ion
%  coupled to {\it itinerant} Nb$^{4+}$ carriers induces 
%  a new two dimensional Kondo-Heisenberg lattice system. 
In $3d$ systems, a much stronger correlation strength leads 
to a Mott transition, which has been intensively discussed,\cite{andersen} 
while SOC is usually minor. 
Here, we focus on the recently synthesized \bfpo~ compound with 
a honeycomb lattice of high spin $d^6$ Fe$^{2+}$ ions,   
leading to an effectively isolated minority-spin $d^1$ configuration 
due to a large exchange splitting of the high spin Fe ion. 
Large $3d$ orbital moments
have been encountered recently\cite{shruba}   %% \cite{HB} 
in low-symmetry environments but gaining understanding of their origin is
not simple.

Mentr\'e and coworkers synthesized a rare example of 
an insulating 2D Ising ferromagnetic (FM) oxide \bfpo, 
confirmed by the observed critical exponents
in good agreement with the theoretical Ising values, 
with the Curie temperature $T_C = 65$ K and negligible hysteresis.\cite{bfpo1,bfpo2}
This system also shows an unusual re-entrant structural transition sequence:
a re-entrant rhombohedral ($R\bar{3}$) $\rightarrow$ triclinic ($P\bar{1}$ at 140 K)
$\rightarrow$ rhombohedral ferromagnetic (FM) ($R\bar{3}$ at 70 K).
This rare reentrant sequence was suggested to arise from
competition between Jahn-teller distortion and magnetism.\cite{bfpo1,bfpo2}
The observed Curie-Weiss moment, 6.16 $\mu_B$,
is substantially enhanced from the spin-only value of 4.9 $\mu_B$ for a $S=2$ system. 
Additionally, the saturated moment is about 5 $\mu_B$ 
from powder neutron-diffraction studies.\cite{bfpo1,bfpo2} 
These values are close to those of the case of $S=2$, $L=1$,
thus requiring a large orbital moment on the $3d$ ion.
The latter corresponds to the possible largest orbital moment $M_L=1$ $\mu_B$
for the $t_{2g}^1$ system.\cite{bnoo1} 
Optical spectroscopy measurements show semiconducting behavior
with an estimated energy gap of 1.5 eV.\cite{bfpo3} 

Although the generalized gradient approximation (GGA) plus on-site Coulomb repulsion $U$ 
calculations by Mentr\'e and coworkers provided some preliminary results,\cite{bfpo1}
%dominant FM nearest neighbor exchange parameter $J_1\approx2(\pm0.5)$ meV 
%and negligible second $J_2$ and third neighbors $J_3$ interactions,\cite{bfpo1}
available information about the electronic structures is still limited. 
Here we report more extensive density functional theory (DFT) based studies, 
including correlation and SOC,
to uncover the origin of the strong magnetocrystalline anisotropy, 
Mott transition, and the large orbital moment in \bfpo.

\begin{table}[bt]
\caption{Optimized internal parameters and structure information 
with the experimental lattice parameters 
of $a$=4.869 \AA~ and $c$=23.230 \AA~  at low $T$.\cite{bfpo2}
In the representation of a hexagonal lattice,
the Ba ions are at $3a$ (0,0,0). 
The Fe, O1, and P atoms sit at $6c$ (0,0,$z$),
while O2 lies at $18f$ ($x,y,z$).
The O2 ions are corner shared by the PO$_4$ tetrahedra and the FeO$_6$ octahedra, 
whereas the O1 ions are at one of the vertices of the PO$_4$ tetrahedra.
The bond length $d$ and the bond angles are given in units of angstrom and degree, 
respectively.}
\begin{center}
\begin{tabular}{ccccccc}\hline\hline
% ~ & ~ & \multicolumn{5}{c}{$m_l$} \\\cline{3-7}
 ~ & Fe & P & O1 & \multicolumn{3}{c}{O2} \\\cline{5-7}
   internal &  $z$ & $z$ & $z$ & $x$ & $y$ & $z$ \\
   parameter& 0.1680 & 0.5710 & 0.6371 & 0.3494 & 0.0235 & 0.8840 \\\hline \hline
\multicolumn{2}{l}{$d$(Fe-O)($\times$3)} & \multicolumn{5}{c}{2.042, 2.050}\\
\multicolumn{2}{l}{$d$(P-O)}  & \multicolumn{5}{c}{1.536, 1.598($\times$3)} \\
\multicolumn{2}{l}{\angle O-Fe-O($\times$3)} &\multicolumn{5}{c}{86.6, 88.6, 93.2,91.8}\\ 
\multicolumn{2}{l}{\angle O-P-O ($\times$3)} & \multicolumn{5}{c}{107.2, 111.7}\\\hline\hline      
\end{tabular}
\end{center}
\label{table1}
\end{table}

\section{Crystal Structure and methods}
Figure \ref{str} shows the rhombohedral structure of the space group $R\bar{3}$ (no. 148), 
which is observed in both low and high $T$ regimes.
In this structure, 
P ions sit in the center of the Fe honeycomb lattice, but above and below the layer
along the $\hat{c}$ direction.
The P ions are edge shared with three oxygen ions of the FeO$_6$ octahedra,
and form PO$_4$ tetrahedra.
The Fe$^{2+}$ layers are isolated by the intervening insulating 
(PO$_4$)$^{3-}$ tetrahedra and Ba$^{2+}$ ions.
In the low $T$ phase, the interlayer distance is 7.71 \AA~ along the $\hat{c}$ axis,
while the Fe-Fe distance is 2.81 \AA.
Both structurally and electronically (below), the compound is strongly 2D in nature.

Using the GGA exchange-correlation functional, 
the internal parameters\cite{opt,fplo1} were optimized at the
experiment lattice parameters for the low $T$ phase 
(see Table \ref{table1}),
since no experimental data are available.
The optimized positions are very similar to the experimental ones at high $T$,
as expected since the difference in volumes between the low and high $T$ phases 
is only several tenths of a percent.\cite{bfpo1,bfpo2}
The FeO$_6$ octahedra are somewhat more regular in the low $T$ phase,
while the PO$_4$ tetrahedra are close to an ideal tetrahedron in the high $T$ phase.  

Our calculations were performed in a rhombohedral unit cell
with GGA exchange-correlation functional plus $U$ and SOC, 
which are implemented in 
the accurate all-electron full-potential code {\sc Wien2k},\cite{wien2k}
which uses the tetrahedron method for Brillouin-zone integrations.
The fully localized limit flavor of the GGA+$U$ functional was used for
this strong Fe moment compound.
On-site Coulomb repulsion $U=3-5$ eV and the Hund's exchange parameter $J=0.70$ eV 
were employed, as widely used for Fe$^{2+}$ systems.\cite{srfeo2,bfpo1}
The basis size was determined from the smallest atomic sphere by
$R_{mt}K_{max}=7$ and augmented plane-wave radii
of Ba 2.30, Fe 2.00, P 1.61, and O 1.29, in atomic units.
The Brillouin zone was sampled up to a $k$ mesh of $21\times 21\times 21$
to check convergence carefully in this system, which is nonmetallic at
almost all levels of the theory.

\section{Results}

\begin{figure}[tbp]
%\vskip 8mm
\resizebox{8cm}{6cm}{\includegraphics{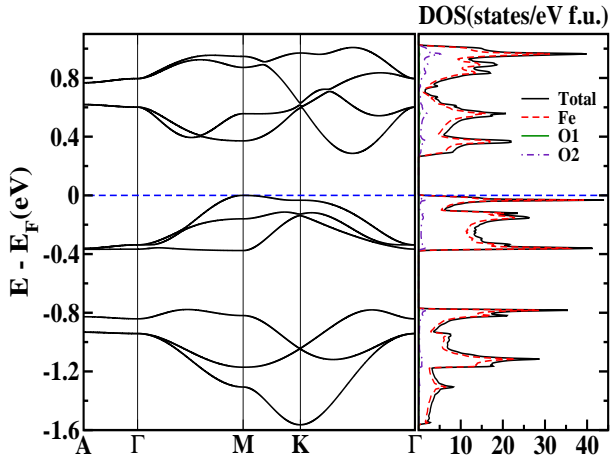}}
\caption{(Color online) Nonmagnetic GGA band structure (left) 
 and orbital projected densities of states (right) of the 
 rhombohedral structure
 in the energy region containing Fe $d$ orbitals, not that the
 O1 contribution is negligible and invisible here.
 The occupation confirms the $d^6$ Fe$^{2+}$ configuration and
 the density of states just below the gap shows a 
 quasi-one-dimensional character due to the flat band that
extends from $M$ to $K$, hence all around the edge of the Brillouin zone.
 The bottom of the gap is denoted 
 by the horizontal dashed line.
The $k$-point labels are for the conventional hexagonal lattice,
viz. graphene.
% The experiment structure\cite{bfpo1} obtained in the high $T$ regime was used.
}
\label{nband}
\end{figure}

\subsection{Underlying electronic structure}
Before discussing the low-temperature FM phase results, we address the nonmagnetic state, 
which represents the underlying electronic structure.
The enlarged band structure in the energy range of Fe $d$ orbitals 
and the corresponding densities of states (DOSs) are given in Fig. \ref{nband}.
In the region displayed there are ten $d$ bands, since the unit cell contains two Fe ions.
From the tiny oxygen character in the DOS around $E_F$,
we conclude that direct Fe $t_{2g}-t_{2g}$ hopping is substantial 
due to the edge-sharing octahedral structure,  
and this interaction leads to a bonding-antibonding splitting $\sim$1 eV 
of the fully filled $t_{2g}$ manifold.
A gap of 0.25 eV opens between the antibonding $t_{2g}^\ast$ state 
and the unfilled $e_g$ manifold.
This feature of isolated bands around $E_F$ in a hexagonal lattice system is similar to 
that of the hole-doped superconductor Li$_{1-x}$NbO$_2$.\cite{linbo2_1,linbo2_2}
In both bonding and antibonding manifolds,
Dirac points appear at the $K$ point by symmetry, 
as observed in several honeycomb lattices.

An anomalous feature is that the top of the $t_{2g}^\ast$ manifold (bottom of the gap) 
is flat along the $M-K$ line, hence along the {\it entire edge} of the Brillouin zone
(which consists of alternating $M-K$ and $K-M$ lines).
Considering also the dispersionless character along the $\hat{c}$ axis 
due to the large interlayer distance,
the electronic structure becomes an unusual quasi-one-dimensional one, 
as evident in the DOS near $E_F$.
This one-dimensionality, however, is not uniaxial but instead radial, perpendicular
to the edge of the Brillouin zone. A similar one-dimensionality was found as well
in the hexagonal lattice CuAlO$_3$ compound.\cite{cualo3}

\begin{figure}[tbp]
%\vskip 8mm
{\resizebox{7cm}{8cm}{\includegraphics{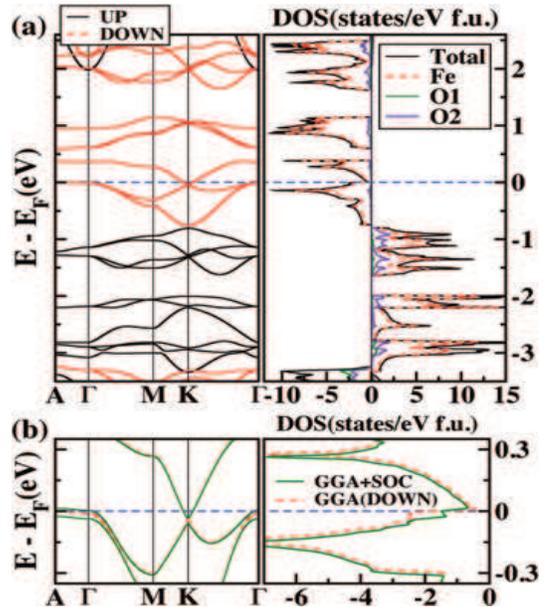}}}
\caption{(Color online) (a) FM band structures and DOSs within GGA 
 in the range of Fe $d$ bands, indicating a half-semimetallic character
 in the minority bands (red dashed lines).
 A Dirac point appears almost exactly at $E_F$ at the point $K$.
%at $\Gamma$.
 (b) Overlaid FM GGA (red dashed lines) and GGA+SOC (green solid lines) 
 band structures and total DOS enlarged near $E_F$, in a $\pm$0.3-eV region 
  containing only some of the minority $t_{2g}$ manifold.
 The Dirac point is split by about 30 meV by SOC.
 Note that a tiny hole pocket compensating a tiny electron pocket at the $K$ point 
 appears in a region that is not visible along the lines shown.
}
\label{fband}
\end{figure}

\begin{figure}[tbp]
%\vskip 8mm
{\resizebox{7.7cm}{12cm}{\includegraphics{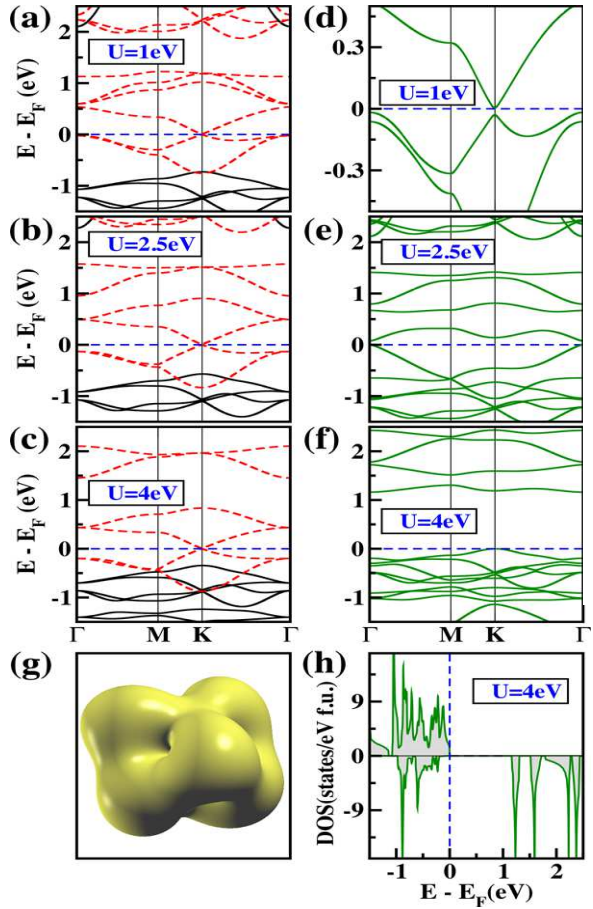}}}
\caption{(Color online) Band structures near E$_F$: (a)-(c) within GGA+$U$,
 and (d)-(f) from GGA+$U$+SOC,
 for $U=1-4$ eV and $J=0.7$ eV.
The spin-up and -down bands in (a)-(c) are denoted by black solid and red dashed lines, respectively,
 while the bands including SOC (d)-(f) are shown by green solid lines.
In (d) the scale is enlarged to allow the gap opening at the Dirac
point to be seen. 
 (g) Charge density plot of the minority $d$ orbitals, shown at the isovalue of 0.09 $e$/\AA$^3$.
 The $c$ axis is along the vertical direction.
 This shape is nearly identical for all calculations performed here.  
% is the $|\phi_{\pm1}|^2$ character.
 (h) total DOS at $U=4$ eV in GGA+SOC+$U$, 
 showing a 1.2-eV energy gap between the corresponding up and down bands.  
 %At $U=1$ eV, the Dirac-point pinpoints $E_F$, 
 %insensitive to a selection of $J$ in the range of 0.5--1 eV, as shown in (a).
}
\label{uband}
\end{figure}

\begin{figure}[tbp]
%\vskip 8mm
%\rotatebox{-90}{\resizebox{6cm}{8cm}{\includegraphics{Fig5.eps}}}
\resizebox{8cm}{6cm}{\includegraphics{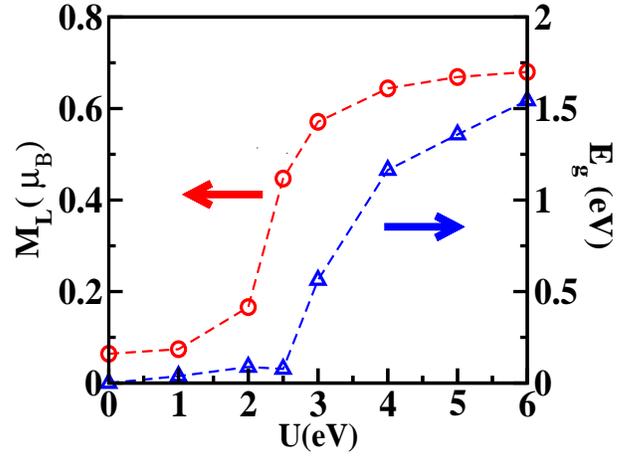}}
\caption{(Color online) Variation of the Fe orbital $M_L$ moment, 
 and energy gap $E_g$, 
 varying $U$ in GGA+$U$+SOC from 0 to 6 eV, with the spin directed along the $\hat{c}$ axis.
 From $U=4$ eV, the gap is between the filled spin-up and the unfilled spin-down channels.
 The critical value of $U$ is in the 2.5--3-eV range.
}
\label{mom}
\end{figure}

\subsection{Ferromagnetic electronic structure}

\subsubsection{Uncorrelated electronic structure and magnetism}
Now we focus on the FM state, which is the ground state at the low $T$ regime.
The magnetic energy is large, with the FM state being 1.2 eV/Fe lower in energy than
for the nonmagnetic state. 
This very large magnetization energy guarantees that the local moment is robust
even in the high $T$ disordered regime.
Based on the Stoner picture (which may be accurate
only for small moments) where this energy is $\frac{1}{4}IM_S^2$, $M_S$ =
4$\mu_B$ leads to a broad first estimate of the Stoner $I$ of 0.3 eV. 
A more realistic means of estimating $I$ is from the band exchange
splitting  
$IM_S$ = $\Delta_{ex}$ = 3 eV of the GGA bands, giving $I$ = 0.75 eV,
more in line with accepted values for the magnetic $3d$ ions.

The energies of FM alignment and in-plane antiferromagnetic (AFM) alignment
have been compared, varying the value of the on-site repulsion $U$. For $U$
equal to 3 eV or less, FM alignment is favored. For $U$=4-5 eV, however, AFM
alignment is favored. In this range the orbital occupations for FM alignment
do not change, whereas there is some small evolution of orbital occupations
for AFM order. The orbital moment is the same large value in all cases.
Evidently this question of magnetic alignment
is near a delicate balance,
so we have not pursued this question of trying to obtain the
nearest-neighbor exchange coupling either in plane (which this difference
reflects), or out of plane, where coupling should be much smaller and also
involves the difficult question of frustrated coupling. We proceed with
analysis using the observed FM alignment.

The left side of Fig. \ref{fband}(a) displays the GGA FM band structure,
showing a half-semimetallic character with a minority-spin 
Dirac point (or Weyl node)\cite{wan}
lying right below $E_F$ by 30 meV at the point $K$.
The corresponding DOS, given in the right side of Fig. \ref{fband}(a),
shows a $\frac{1}{3}$-filled minority $t_{2g}$ manifold,
confirming the high spin $S$ = 2 Fe$^{2+}$ configuration.
The $t_{2g}$ manifold shows a large exchange splitting of 3 eV.
All majority states ($t_{2g}$ and $e_g$) are filled, hence $\frac{5}{2}$ spin,
and orbitally inert. Henceforward we focus on the single minority
electron, whose spin half reduces the total spin to $S$ = 2 and
introduces the question of orbital occupation.

With the local $X,Y,Z$ coordinates of the FeO$_6$ octahedron, 
the $t_{2g}$ orbitals with global $z$ axis along $\hat c$ can be
described in a three-fold symmetry adapted manner by
\begin{eqnarray}
\phi_m=\frac{1}{\sqrt{3}}(\xi_m^0d_{XY}+\xi_m^1d_{YZ}+\xi_m^2d_{ZX}),
\label{eq1}
\end{eqnarray}
where the phase factor is $\xi_m=exp(i\frac{2\pi m}{3})$, $m=L_z$
is the projection of the orbital moment, and the superscript is
an exponent.
Due to the small trigonal distortion, the $L_z=0$ orbital $\phi_0$ ($d_{z^2}$
in shape) has 
a somewhat higher site energy than the doublet $\phi_{\pm1}$.
The corresponding DOS near $E_F$ has some similarity to a particle-hole
asymmetric version of graphene,\cite{nandk} due to a Dirac point at $K$
that lies very near the Fermi level.
The strong magnetism completely changes the Fe ion configuration
from \{$t_{2g}^6 e_g^0$\} to \{$t_{2g}^{3\uparrow} e_g^{2\uparrow} t_{2g}^{1\downarrow}$\}.

\subsubsection{Dirac point: Semimetal-insulator transition} 
We now address the observed insulating character of \bfpo, using the
GGA+$U$ approach for correlated insulators.
Applying a small $U=1$ eV with $J=0.5-1$ eV to the Fe ions within GGA+$U$,
the Dirac point degeneracy pins $E_F$ and no gap opens, as shown in Fig. \ref{uband}(a).
The Dirac cone is isotropic, as in graphene.
As $U$ is increased, the Dirac point degeneracy continues to pin E$_F$.
Applying $U$ up to 7 eV, beyond the limit of a proper value for Fe ions,
although higher energy bands shift this system remains a half-semimetal without gap, 
as illustrated in Fig. \ref{uband}(b) and (c).
%  Note that a band with mostly $\phi_{+1}$ character shifts toward $E_F$
%  from the $t_{2g}^\ast$ manifold as $U$ increases (see below). 

As mentioned in the Introduction, experiment shows
semiconducting behavior and implies a large orbital moment $M_L$,\cite{bfpo1,bfpo2} 
which are not explained by GGA+$U$.
We carried out calculations including SOC, both GGA+SOC, and GGA+$U$+SOC.
The GGA+SOC band structure, overlapping the GGA one,
is shown in the left panel of Fig. \ref{fband}(b).
SOC leads to a band splitting of 40 meV near $E_F$, 
visible in this plot only at the $\Gamma$ and $K$ points.
The linear Dirac bands are also split by 30 meV.
This small effect is consistent with the small SOC strength
in $3d$ ions, and spin mixing is also degraded by the 2.5-eV
exchange splitting.

Another interesting feature occurs in the parabolic bands touching $E_F$ 
at the $\Gamma$ point.
Inclusion of SOC splits the bands, degenerate at the $\Gamma$ points,
and leaves only one maximum at $E_F$, which is more isotropic 
than in the case excluding SOC.
As a result, as shown in the DOS of the right side of Fig. \ref{fband}(b),
a van Hove singularity (vHs) 
appears very near $E_F$,
in addition to two additional vHs at --0.15 and 0.25 eV.

Simultaneous inclusion of correlation and SOC effects 
splits the Dirac point degeneracy, thereby opening a gap 
for $U$ as small as 1 eV. At small $U$ this gap is a spin-orbit gap, not a Mott gap.
Although the strength of SOC is small in $3d$ systems,
the symmetry lowering due to SOC becomes crucial for opening a gap in this system,
as occurs also in BaCrO$_3$.\cite{bacro3}
The resulting band structures in GGA+$U$+SOC are given in Fig. \ref{uband}(d)-(f).
As displayed in Fig. \ref{mom}, the energy gap of 1.2 eV for $U=4$ eV is close to 
the experimentally estimated value.
% At $U=4$ eV, the top of the filled bands has the spin-down character at the $\Gamma$-point 
% and the spin-down character around the $K$-point.
The gap is indirect, between $K$ and $\Gamma$, although due to the
small dispersion the direct gap is not much larger. 
There has been some uncertainty in determining the optical gap 
spectroscopically.\cite{bfpo3}

\subsubsection{Degeneracy lifting and large orbital moment}
GGA+SOC gives a negligible magnetocrystalline anisotropy,
with an orbital moment of $M_L\approx0.08 \mu_B$/Fe both for in-plane and 
$\hat c$-axis spin alignment.  
Varying $U$ in GGA+$U$+SOC with the spin along $\hat c$, 
$M_L$ of Fe rapidly increases in the $U$=2--3-eV range, then saturates
near 0.7$\mu_B$ at $U=4$ eV, as displayed in Fig. \ref{mom}.
This value excludes contributions from the tails of the $3d$
orbitals extending out of the Fe sphere, so the atomic value will
be somewhat larger. (This large orbital moment arises for antiferromagnetic
alignment as well.) As a critical value $U_c$ of 2.5--3 eV, the
opening of a Mott gap begins.

Consistent with the large exchange splitting $\Delta_{ex}$ = 3 eV, 
the total spin moment $M_S=4 \mu_B$/Fe remains unchanged,
since mixing of the spin-split orbitals by small SOC is minor.
At $U=4$ eV, the spin direction along the $\hat c$ direction has much lower
energy around 400 meV/f.u. than for spin in the $\hat a-\hat b$ plane,
explaining the observed Ising behavior.
The state obtained for the spin direction in the $\hat a-\hat b$ plane (not pictured)
remains semimetallic (Dirac point) even in the large $U$ regime, showing a very small 
$M_L$ of $\sim$0.06 $\mu_B$.
This striking anisotropy  confirms that SOC drives the semimetal-insulator
transition, and that depends strongly on spin direction.
% The Dirac point remains unsplit in the spin direction.
This type of Ising magnet, i.e., with the spin direction perpendicular to the layers,
has been proposed as a crucial ingredient 
for emergence of a quantum anomalous Hall state\cite{zhou} in ferromagnets.  

We now discuss the origin of the large orbital moment. The SOC operator is
\begin{eqnarray}
H_{SO} = \xi \vec S \cdot \vec L = \xi [ S_z L_z +\frac{1}{2}(S_+L_- + S_- L_+)].
\end{eqnarray}
The majority shell is filled, spin split from the active minority orbital by
$\Delta_{ex}$, and inert, because spin mixing is reduced by $\xi/\Delta_{ex} \sim$10$^{-2}$.
Thus the spin and orbital operators are those of the minority channel. With
the majority channel frozen out, the only effect is from the diagonal term
$S_z L_z$ with $S_z = -\frac{1}{2}$. For the magnetization along the $\hat c$ axis,
the orbital degeneracy is broken by $-\frac{\xi}{2}L_z$ for the two orbital
projections. With a bandwidth of 0.5 eV, this small splitting induces an orbital
moment, but more importantly it specifies the orbital to be occupied in the
Mott insulating phase. For the magnetization in plane, the $L_z = \pm1$
degeneracy remains unbroken, no orbital moment arises, and correlation is ineffective
in opening a gap, as discussed above.

As can be seen in Fig. \ref{uband}(g),
the spin-density plot of the filled minority $t_{2g}$ manifold shows
the $\phi_{+1}$ character characteristic of an $L_z=1$ orbital.
When symmetry allows it, the two orbitals $\phi_{\pm 1}$ are equally 
filled, the semi-Dirac degeneracy remains at $K$, and so no gap emerges. It
is SOC that breaks the degeneracy, which though small, enables
the correlation effects to operate.
Upon increasing $U$, the single minority electron density begins to transfer
from the $\phi_{-1}$ orbital to the $\phi_{+1}$ orbital.

\section{Discussion and Summary}
Using the DFT-based calculations, we have investigated
the Ising insulating Fe honeycomb lattice system \bfpo, 
which at the GGA level results in
a half-semimetallic electronic structure with a fermionic Dirac point
lying nearly precisely at $E_F$; this degeneracy is split by inclusion of SOC.
Applying both $U$ and SOC suppresses the direct $t_{2g}-t_{2g}$ interaction,
and leads to a Mott transition of uncommon origin, becoming enabled by
the symmetry lowering due to the (small in magnitude) SOC.
In the insulating phase the magnetic phase is
$S=2$, $L=1$. The atomic-like value of $L$ suggests classifying \bfpo~as
a $J_{eff}=3$ Dirac Mott insulator.
These DFT-based studies have explained the large magnetic anisotropy,
the very large orbital moment for a $3d$ ion, and the unusual manner
in which spin-orbit coupling enables the Mott gap to open in \bfpo.
Even the magnitude of the gap is given reasonably by our methods.

The re-entrant structural transitions mentioned in the Introduction
needs comment. The Fe honeycomb lattice studied here bears a
great deal of similarity to the (111)-oriented perovskite bilayers
that are recently being studied,\cite{doennig,doennig2} a difference
being that in the latter case the metal-oxygen octahedra are corner
sharing rather than edge sharing. In those systems structural
symmetry breaking is predicted in some cases, using the same methods
as used here. That symmetry breaking involves an unusual
type of orbital-occupation competition. The high-symmetry structure
(threefold, as here)
supports complex-valued, orbital moment containing
orbitals, but breaking to low symmetry ($P1$ space group in that case)
is connected with occupation of $t_{2g}$
type real orbitals adapted to the local pseudocubic symmetry. 

The same orbital-occupation competition can be expected in \bfpo.
At high temperature where the Fe spins are disordered (primarily
up and down due to the Ising nature) the (space- and time-averaged)
symmetry is the high threefold symmetry, and the symmetry adapted
$\phi_{+1}$ orbital is occupied and shows up in the Curie-Weiss
moment. The ``ground state'' (with
spins still disordered) may be one
in which a real valued $t_{2g}$ orbital is ordered instead, with
accompanying symmetry lowering, and
the transition toward such a lower-symmetry structure occurs at 140 K. 
However, upon ordering ferromagnetically, the strains that accompany occupation
of cubic orbitals (viz. Jahn-Teller strains) may be incompatible
with long-range order. Such spin-lattice competition could then favor a return to
the symmetric structure at or near the Curie temperature, as
observed. Further consideration of this mechanism will be left
for future work. 

A Dirac point has played an important role in our discussion of how
the Mott gap in \bfpo~arises only when spin-orbit coupling is included.
Few instances of a Dirac point pinned to the Fermi level
in a ferromagnet have been reported, which
produces a half-semimetallic system.
One such instance is in the  
CrO$_2$/TiO$_2$ multilayer\cite{cai}, where it is actually a
semi-Dirac semi-Weyl point.  Another instance is that of
(111)-oriented perovskite superlattices,\cite{doennig,cook} where Dirac point
degeneracies are only lifted by symmetry-breaking interactions.
Related band points arise in the hexagonal W lattice on the Cl-Si surface,\cite{zhou}
and are also of interest in triangular lattice systems.\cite{akagi}
In some such systems, Chern bands with a Chern number $\pm$2 are proposed.\cite{cai,cook}
In addition, the three-dimensional Weyl semimetals NbP and TaAs, 
which are distinguished by lack of inversion symmetry but 
preserving time reversal-symmetry, 
have been very much of interest, since these compounds show 
an extremely large and unsaturated magnetoresistivity 
even for very high magnetic field.\cite{huang,weng,NbP,NbP2}
\bfpo~ bears some similarity to the above-mentioned compounds, and 
has the potential for additional interesting properties.
%In particular, at $T\sim55$ K, below $T_C$, 
%a kink in the magnetic moment along the $c$-axis,
%measured by the power neutron diffraction,\cite{bfpo2} may indicate
%other undetected transition in \bfpo.
%It suggests further careful experimental research on the system. 

\begin{acknowledgments}
 We acknowledge S. Y. Savrasov for a useful discussion on the Dirac point,
 R. Pentcheva for discussions of (111) transition-metal bilayers, and 
 A. S. Botana for technical discussions. 
 This research was supported by National Research Foundation of Korea 
 Grant No. NRF-2013R1A1A2A10008946 (Y.J.S. and K.W.L.),
 by U.S. National Science Foundation Grant No. DMR-1207622-0 (K.W.L.)
 and by U.S. Department of Energy Grant No. DE-FG02-04ER46111 (W.E.P.).
\end{acknowledgments}


\begin{thebibliography}{10}
%TI
\bibitem{kane} C. L. Kane and E. J. Mele,
 $Z_2$ Topological Order and the Quantum Spin Hall Effect.
 Phys. Rev. Lett. {\bf 95}, 146802 (2005).

%quasi-honeycomb
\bibitem{fiete} A. R\"{u}egg and G. A. Fiete,
 Topological insulators from complex orbital order 
 in transition-metal oxides heterostructures.
 Phys. Rev. B {\bf 84}, 201103(R) (2011).

\bibitem{shitade} A. Shitade, H. Katsura, J. Kune\v{s}, X.-L. Qi, 
 S.-C. Zhang, and N. Nagaosa,
 Quantum Spin Hall Effect in a Transition Metal Oxide Na$_2$IrO$_3$.
 Phys. Rev. Lett. {\bf 102}, 256403 (2009).

%Dirac half-metal 1
\bibitem{zhou} M. Zhou, Z. Liu, W. Ming, Z. Wang, and F. Liu,
 $sd^2$ Graphene: Kagome Band in a Hexagonal Lattice.
 Phys. Rev. Lett. {\bf 113}, 236802 (2014). 

%d^1 system
\bibitem{bnoo1} K.-W. Lee and W. E. Pickett,
 Orbital-quenching-induced magnetism in Ba$_2$NaOsO$_6$.
 Europhys. Lett. {\bf 80}, 37008 (2007).

\bibitem{bnoo2} S. Gangopadhyay and W. E. Pickett,
 Spin-orbit coupling, strong correlation, and insulator-metal transitions: 
 The J$_{eff}$=3/2 ferromagnetic Dirac-Mott insulator Ba$_2$NaOsO$_6$.
 Phys. Rev. B {\bf 91}, 045133 (2015).

\bibitem{jyu} H. Jin, H. Jeong, T. Ozaki, and J. Yu,
 Anisotropic exchange interactions of spin-orbit-integrated states in Sr$_2$IrO$_4$.
 Phys. Rev. B {\bf 80}, 075112 (2009).

% \bibitem{nb12} K.-W. Lee and W. E. Pickett,
%  Organometalliclike localization of $4d$-derived spins in an inorganic conducting 
%  niobium suboxide.
%  Phys. Rev. B {\bf 91}, 195152 (2015).

\bibitem{andersen} E. Pavarini, S. Biermann, A. Poteryaev, A. I. Lichtenstein, 
 A. Georges, and O. K. Andersen,
 Mott transition and suppression of orbital fluctuations 
 in orthorhombic $3d^1$ perovskites.
 Phys. Rev. Lett. {\bf 92}, 176403 (2004).

\bibitem{shruba} I. G. Rau, S. Baumann, S. Rusponi, F. Donati, S. Stepanow, 
 L. Gragnaniello, J. Dreiser, C. Piamonteze, F. Nolting, S. Gangopadhyay, 
 O. R. Albertini, R. M. Macfarlane, C. P. Lutz, B. A. Jones, P. Gambardella, 
 A. J. Heinrich, and H. Brune,
 Reaching the magnetic anisotropy limit of a $3d$ metal atom.
 Science {\bf 344}, 988 (2014).

% \bibitem{HB}H. B. Rhee and W. E. Pickett, 
%   Strong interactions, narrow bands, and dominant spin-orbit coupling in Mott 
%   insulating quadruple perovskite CaCo$_3$V$_4$O$_{12}$,
%   Phys. Rev. B {\bf 90}, 205119 (2014).

%exp
\bibitem{bfpo1} H. Kabbour, R. David, A. Pautrat, H.-J. Koo, M.-H. Whangbo, 
 G. Andr\'{e}, and O. Mentr\'{e},
 A genuine two-dimensional Ising ferromagnet with magnetically driven re-entrant transition.
 Angew. Chem. Int. Ed. {\bf 51}, 11745 (2012).

\bibitem{bfpo2} R. David, A. Pautrat, D. Filimonov, H. Kabbour, H. Vezin,
 M.-H. Whangbo, and O. Mentr\'{e},
 Across the structural re-entrant transition in \bfpo: influence 
 of the two-dimensional ferromagnetism.
 J. Am. Chem. SOC. {\bf 135}, 13023 (2013).

\bibitem{bfpo3} R. David, H. Kabbour, D. Filimonov, M. Huv\'{e},
 A. Pautrat, and O. Mentr\'{e},
 Reversible topochemical exsolution of iron in BaFe$^{2+}_2$(PO$_4$)$_2$.
 Angew. Chem. Int. Ed. {\bf 53}, 13365 (2014).


%methods
\bibitem{opt} Using {\sc fplo},\cite{fplo1}
the internal parameters were optimized, until the forces were less than 1 meV/a.u.

\bibitem{fplo1} K. Koepernik and H. Eschrig, 
 Full-potential nonorthogonal local-orbital minimum-basis band-structure scheme.
 Phys. Rev. B {\bf 59}, 1743 (1999).

\bibitem{wien2k} K. Schwarz and P. Blaha,
 Solid state calculations using WIEN2k.
 Comput. Mater. Sci. {\bf 28}, 259 (2003).


% U, J for Fe2+
\bibitem{srfeo2} H. J. Xiang, S.-H. Wei, and M.-H. Whangbo,
 Origin of the structural and magnetic anomalies of the layered compound SrFeO$_2$: 
 a density functional investigation.
 Phys. Rev. Lett. {\bf 100}, 167207 (2008).

%I for Fe
%\bibitem{gunnarsson} O. Gunnarsson,
% Band model for magnetism of transition metals in the spin-density-funational 
% formalism.
% J. Phys. F: Metal Phys. {\bf 6}, 587 (1976). 

%Extended vHS
%\bibitem{gofron} K. Gofron, J. C. Campuzano, A. A. Abrikosov, M. Lindroos, 
% A. Bansil, H. Ding, D. Koelling, and B. Dabrowski,
% Observation of an "Extended" Van Hove Singularity in YBa$_2$Cu$_4$O$_8$ 
% by Ultrahigh Energy Resolution Angle-Resolved Photoemission.
% Phys. Rev. Lett. {\bf 73}, 3302 (1994).

%LiNbO2
\bibitem{linbo2_1} E. R. Ylvisaker and W. E. Pickett,
 First-principles study of the electronic and vibrational properties of LiNbO$_2$.
 Phys. Rev. B {\bf 74}, 075104 (2006).

\bibitem{linbo2_2} K.-W. Lee, J. Kune\v{s}, R. T. Scalettar, and W. E. Pickett,
 Correlation effects in the triangular lattice single-band system Li$_x$NbO$_2$.
 Phys. Rev. B {\bf 76}, 144513 (2007).

%quasi-1D electronic structure
\bibitem{cualo3}E. R. Ylvisaker and W. E. Pickett,
  Doping-induced spectral shifts in two-dimensional metal oxides.
  EPL (Europhys. Lett.) 101, 57006 (2013).  


%FSM
%\bibitem{fsm} K. Schwarz and P. Mohn, 
% J. Phys. F:Met. Phys. {\bf 14}, L129 (1984).

%Weyl semimetal1
\bibitem{wan} X. Wan, A. M. Turner, A. Vishwanath, and S. Y. Savrasov,
 Topological semimetal and Fermi-arc surface states in the electronic structure of 
 pyrochlore iridates.
 Phys. Rev. B {\bf 83}, 205101 (2011).

%hexagonal lattice
\bibitem{nandk} R. Nandkishore, L. S. Levitov, and A. V. Chubukov,
 Chiral superconductivity from repulsive interactions in doped graphene.
  Nature Phys. {\bf 8}, 158 (2012).

%BaCrO3
\bibitem{bacro3} H.-S. Jin, K.-H. Ahn, M.-C. Jung, and K.-W. Lee, 
Strain and spin-orbit coupling induced orbital ordering in the Mott insulator BaCrO$_3$.
Phys. Rev. B {\bf 90}, 205124 (2014).

%Weyl semimetal2
\bibitem{doennig2} D. Doennig, W. E. Pickett, and R. Pentcheva,
 Massive Symmetry Breaking in LaAlO$_3$/SrTiO$_3$(111) Quantum Wells: A 
 Three-Orbital, Strongly Correlated Generalization of Graphene.
 Phys. Rev. Lett. {\bf 111}, 126804 (2013).

\bibitem{doennig} D. Doennig, W. E. Pickett, and R. Pentcheva,
 Confinement-driven transitions between topological and Mott phases 
 in (LaNiO$_3$)$_N$/(LaAlO$_3$)$_M$(111) superlattices.
 Phys. Rev. B {\bf 89}, 121110(R) (2014).

%half-metallic Dirac-point 2
\bibitem{cai} T.-Y. Cai, X. Li, F. Wang, J. Sheng, J. Feng, and C.-D. Gong,
Emergent topological and half semimetallic Dirac Fermions at oxide interfaces.
arXiv:1310.2471.

\bibitem{cook} A. M. Cook and A. Paramekanti,
 Double Perovskite Heterostructures: Magnetism, Chern Bands, and Chern Insulators.
 Phys. Rev. Lett. {\bf 113}, 077203 (2014).

\bibitem{akagi} Y. Akagi and Y. Motome,
 Spontaneous formation of kagome network and Dirac half-semimetal
 on a triangular lattice.
 Phys. Rev. B {\bf 91}, 155132 (2015).


%Weyl
\bibitem{huang} S.-M. Huang, S.-Y. Xu, I. Belopolski, C.-C. Lee, G. Chang,
 B. Wang, N. Alidoust, G. Bian, M. Neupane, A. Bansil, H. Lin, and M. Z. Hasan,
 An inversion breaking Weyl semimetal state in the TaAs material class.
 Nature Commun. {\bf 6}, 7373 (2015).

\bibitem{weng} H. Weng, C. Fang, Z. Fang, B. Andrei Bernevig, and X. Dai,
 Weyl semimetal phase in noncentrosymmetric transition metal monophosphildes.
 Phys. Rev. X {\bf 5}, 011029 (2015).

\bibitem{NbP} C. Shekhar, A. K. Nayak, Y. Sun, M. Schmidt, M. Nicklas, I. Leermakers,
 U. Zeitler, W. Schnelle, J. Grin, C. Felser, and B. Yan,
 Extremely large magnetoresistance and ultrahigh mobility
 in the topological Weyl semimetal NbP.
 Nature Phys. {\bf 11}, 645 (2015). 

\bibitem{NbP2}K.-H. Ahn, K.-W. Lee, and W. E. Pickett,
  Spin-orbit interaction driven collective electron-hole excitations in a 
  non-centrosymmetric nodal loop Weyl semimetal.
  arXiv:1507.03637. 
 


\end{thebibliography}
\end{document}